\documentclass{aa}  
\usepackage{graphicx}
\usepackage{txfonts}
\usepackage{lipsum}
\usepackage{subcaption}         
\usepackage{lscape}             
\usepackage{placeins}           
                                
\usepackage{hyperref}
\hypersetup{
    colorlinks=true,
    citecolor=blue,
    linkcolor=blue,
    filecolor=blue,      
    urlcolor=blue}

\begin{document}

   \title{Living near filaments: How environment shapes age profiles in low-mass galaxies}
   \titlerunning{How environment shapes age profiles}


%
%
%

   \author{Elisa A. Tau\inst{1,2}\corrauth{elisa.tau@usm.cl}        
        \and Diego Pallero\inst{1,2}
        \and Rory Smith\inst{1,2}
        \and Raúl Baier-Soto\inst{1,2}
        \and Antonela Monachesi\inst{3}
        \and Silvio Varela-Lavin\inst{1,2}
        }

   \institute{Departamento de Física, Universidad Técnica Federico Santa María, Avenida España 1680, Valparaíso, Chile
   \and Núcleo Milenio de Galaxias (MINGAL), Chile
   \and Departamento de Astronomía, Universidad de La Serena, Av. Raúl Bitrán 1305, La Serena, Chile
   }

   \date{XXXX, 2026}

 
  \abstract{U-shaped radial age profiles have recently been reported in low-mass galaxies. This type of profile differs from the usually observed positive radial age gradients of the low-mass regime. 
   We aim to characterise the environment in which low-mass galaxies with U-shaped age profiles are more commonly found. 
   We use the IllustrisTNG-50 hydrodynamical simulations to select a sample of galaxies with stellar masses ranging from $10^9 \, M_\odot$ to $10^{10} \, M_\odot$. Based on their radial age profiles, we classify the sample into galaxies with and without a U-shaped feature. We define a U-shaped profile as one where stellar ages decrease towards a local minimum and subsequently increase, requiring an age contrast of at least 0.4 Gyr on either side of the minimum relative to its value. 
   We find 145 galaxies with U-shaped age profiles ($\sim 11\%$ of the sample). They are mostly central systems ($\sim 72\%$) and $\sim 82\%$ of them reside within low-mass halos ($M_{200} < 10^{12} \, M_\odot$). 
   We also find that galaxies exhibiting this feature are located at large distances from galaxy groups but close to cosmic filaments, with a median distance of $\sim 265$ kpc, whereas those without U-shaped age profiles tend to be located farther away from these structures ($\sim 351$ kpc).}


   \maketitle
   \nolinenumbers

\section{Introduction}

Low-mass galaxies exhibiting U-shaped age profiles have recently been reported in both observational and theoretical works \citep[e.g.][]{Pessa2023, Cohen2024, Tau2026}. This kind of profile differs from the predominantly positive radial age gradients commonly reported in low-mass galaxies, where stellar populations become progressively older with increasing galactocentric radius \citep[e.g.][]{Gallart2008, BenitezLlambay2016, Graus2019, Zhuang2021, Liao2023}. In contrast, U-shaped profiles are characterised by a decrease in stellar age towards a certain radius, followed by an increase in age at larger radii, indicating a reversal of the conventional radial age trend in the outskirts. 
Recent zoom-in simulation studies attributed this feature to an interplay between the star formation and the merger histories of isolated systems \citep[][Tau et al., in preparation]{Tau2026}. However, these works focused exclusively on isolated dwarf galaxies.

To fully understand why low-mass galaxies develop a U-shaped age profile, it is crucial to examine the role of the surrounding large-scale environment. 
To date, studies of galaxies with U-shaped age profiles have primarily focused on characterising the shape of these profiles and identifying the intrinsic physical mechanisms responsible for their formation \citep[e.g.][]{Chen2026, Tau2026}, i.e. how the evolutionary history of a galaxy gives rise to the observed radial age distribution. However, the possible role of the surrounding large-scale and local environment remains less explored. In particular, understanding whether U-shaped age profiles are preferentially associated with specific environments, such as proximity to cosmic filaments or particular host halo masses, could provide important clues about the processes that contribute to their formation. Dense filamentary and halo environments host high concentrations of gas and dark matter \citep[e.g.][]{Cautun2014, Kuchner2020} that drive external quenching via ram-pressure stripping \citep{BenitezLlambay2013}, and elevated merger rates \citep{Dulcien2026}, both capable of leaving distinct signatures on age gradients. Investigating the environmental properties of these systems is therefore a necessary step towards establishing whether their U-shaped radial age profiles arise solely from internal evolutionary processes, as presented in previous works \citep[e.g.][]{Tau2026}, or are also influenced by the environments in which they reside.

In this letter, we characterise the environment where low-mass galaxies with U-shaped age profiles are embedded, and compare them to those of low-mass systems without this feature in their age profiles. This letter is structured as follows: in Sec. \ref{sec:methodology} we present the IllustrisTNG simulations and our sample selection, and set a definition for U-shaped profiles; in Sec. \ref{sec:results} we present our results, and we discuss them in Sec. \ref{sec:discussion}; finally, we summarise our findings in Sec. \ref{sec:conclusions}.

\section{Methodology} \label{sec:methodology}

\subsection{IllustrisTNG simulations and sample selection} \label{subsec:tng50}

In this work, we use the magneto-hydrodynamical cosmological IllustrisTNG suite of simulations \citep[TNG,][]{Nelson2019}, which was run with the AREPO code \citep{Springel2010, Pakmor2014}. This simulation adopts a $\Lambda$CDM cosmology consistent with Planck results ($\Omega_m = 0.3089$, $\Omega_b = 0.0486$, $\Omega_\Lambda = 0.6911$, and $h = 0.6774$; \citealt{PlanckCollaboration2016}), and includes a comprehensive model for galaxy formation physics, which is able to realistically follow the formation and evolution of galaxies through time \citep{Weinberger2017, Pillepich2018a}. In particular, we focus on TNG50, which allows us to resolve low-mass galaxies with a good resolution.

TNG50 has baryonic and dark matter particle masses of $m_{\rm bar} = 5.7 \times 10^{4} \,  M_\odot$ and $m_{\rm DM} = 3.1 \times 10^{5} \, M_\odot$, respectively. This simulated box of $\sim 50 \, \rm Mpc^3$ reproduces key observables, including galaxy colour bimodality, the mass-metallicity relation, and galaxy sizes \citep[e.g.][]{Nelson2018, Pillepich2018b, Torrey2018}. Each star particle is treated as a simple stellar population, and information regarding their properties is provided in public catalogues.

Throughout this work, we consider subhalos with a stellar mass ranging from $10^9 \, M_\odot$ to $10^{10} \, M_\odot$ in order to have a good resolution to compute the galaxies' radial age profile across an extension of $10 \, R_{hm}$ from their centres, being $R_{hm}$ the stellar half-mass radius of each galaxy. We made sure all galaxies have a cosmological origin by using the \textsc{SubhaloFlag} parameter provided by TNG50, and that no galaxies more massive than $\frac{1}{100} \times M_*$ were located within $8 \,R_{hm}$ of the main galaxy to avoid contamination (see Appendix \ref{appendix} for details). This cut leaves a total of $1301$ low-mass systems.

\subsection{Definition of U-shaped profiles} \label{subsec:Ushape_definition}

In order to classify the galaxies according to whether they exhibit a U-shaped radial age profile or not, we first set a definition for this feature. Once we obtained the radial age profiles by computing the mean stellar ages using concentric spheres, we computed its sliding median window\footnote{To compute the sliding median, each radial bin $j$ is assigned the median value of the mean ages within a local window extending from $j - 1$ to $j + 1$ (width of $N = 3$ bins), ignoring radial bins that contain no stellar particles.} by taking the median age value within a running local radial interval, so as to smooth the profile and minimise its noise in the outer regions due to the simulation's resolution. All bins have a width of $0.3 \, R_{hm}$. We then selected the profiles' data contained within $8 \, R_{hm}$ in order to avoid spurious minima due to the small number of particles in the outermost parts of the galaxies given the resolution limits. 
For each galaxy, we identified the radial bin containing the minimum median stellar age and evaluated the behaviour of the profile on either side of this minimum. 

We quantified the shape of the profile by calculating the fraction of consecutive radial bins for which the median stellar age decreases towards the minimum and increases beyond it. Specifically, we computed the percentage of negative age differences before the minimum ($S_{\rm before}$) and the percentage of positive age differences after the minimum ($S_{\rm after}$), and defined a mean score as:

\begin{equation}
    S_{\rm U} = \frac{S_{\rm before}+S_{\rm after}}{2}.
\end{equation}

\begin{figure}[ht!]
   \centering
   \includegraphics[width=0.95\columnwidth]{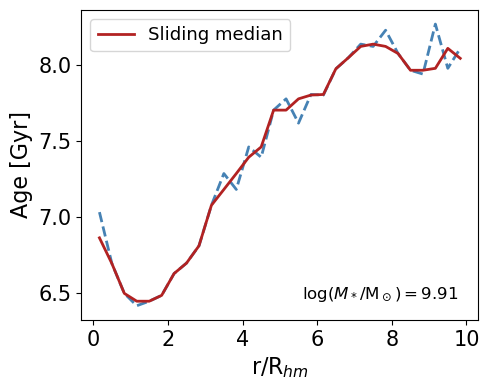}
      \caption{Example of a U-shaped radial age profile (ID=28). The dashed line represents the mean radial age profile computed with concentric spheres. The solid line represents the sliding median window profile computed over the mean age profile.}
         \label{fig:UShapedProfile}
   \end{figure}

\noindent We classified a galaxy as having a U-shaped age profile only if the mean score satisfied that $S_{\rm U} \geq 70\%$. Although a $50\%$ threshold is sufficient to capture this feature, we adopted a stricter $70\%$ cut-off to eliminate spurious classifications. In addition, to ensure that the identified U shape represents a significant variation in stellar age 
rather than a small fluctuation around the minimum, we required an age difference of at least $0.4$ Gyr on each side of the minimum. 
It was shown with the Auriga simulations that a threshold of at least $0.3$ Gyr already allows the identification of the U-shaped feature (Tau et al., in preparation), and it corresponds to age variations sufficiently large to be detected observationally through distinct star-formation tracers like $\text{H}\alpha$ emission \citep[e.g.][]{Tacchella2022}, confirming its robustness for observational testing. Thus, a galaxy's profile was classified as U-shaped when its age predominantly decreased towards a minimum and increased outwards from it, with both sides reaching ages at least 0.4 Gyr older than the minimum. We identified 145 galaxies with U-shaped age profiles ($11.1\%$ of our sample), and we show an example of one in Fig. \ref{fig:UShapedProfile}.

\section{Environmental distribution of galaxies with U-shaped age profiles} \label{sec:results}

There are different approaches to characterise the environment; one is to look at the halo mass in which the galaxies are embedded. We found that $\sim 72\%$ of the galaxies with a U-shaped age profile are central galaxies (105 out of 145), while only $\sim 47\%$ of the galaxies that do not exhibit a U-shaped age profile are central galaxies (546 out of $1156$). We also analysed the halos where these systems reside. As a consequence of this central dominance, galaxies with U-shaped profiles reside overwhelmingly in low-mass halos, with $\sim 82\%$ inhabiting halos of $M_{200} < 10^{12} \, M_\odot$ (and only $\sim 18\%$ satellite systems in $M_{200} \geq 10^{12} \, M_\odot$ halos). In contrast, the higher satellite fraction among systems without this feature results in a broader halo mass distribution ($\sim 53\%$ in $M_{200} < 10^{12} \, M_\odot$ and $\sim 47\%$ in $M_{200} > 10^{12} \, M_\odot$).
We present these results in Fig. \ref{fig:envdist} (also, see Appendix \ref{appendixA}). These statistical trends explain why U-shaped profiles were readily identified in the central low-mass galaxies of the Auriga simulations \citep{Tau2026}.

   \begin{figure}[ht!]
   \centering
    \includegraphics[width=\columnwidth]{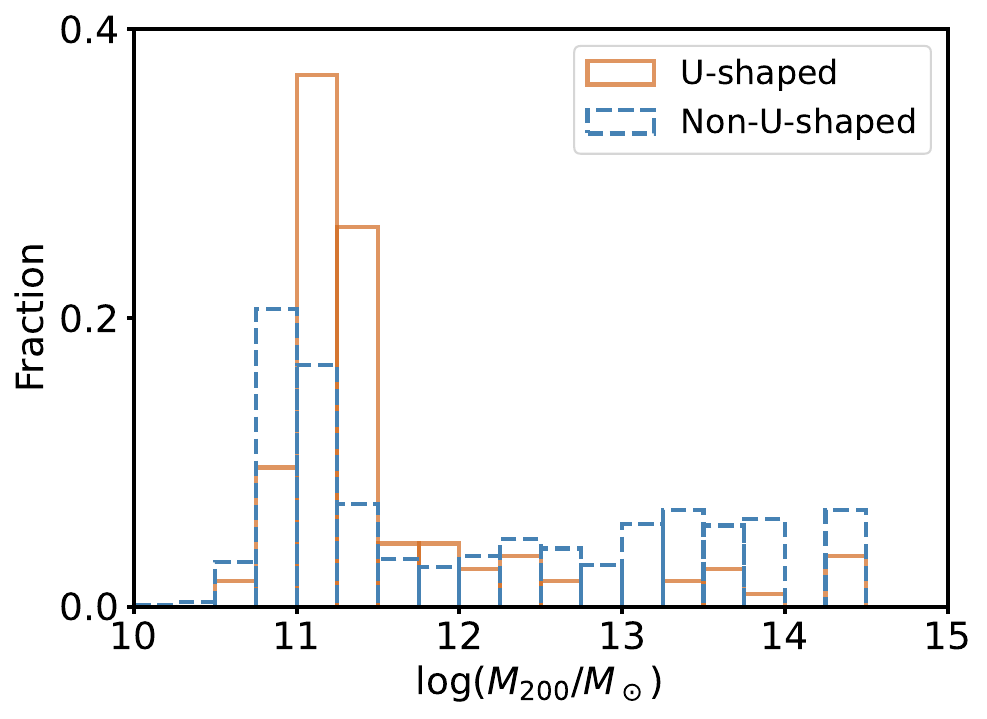}
      \caption{Distribution of the halo mass of the hosts of galaxies with U-shaped age profiles (solid line) and galaxies without these profiles (dashed lines).}
         \label{fig:envdist}
   \end{figure}

   \begin{figure*}[ht!]
   \centering
    \includegraphics[width=2\columnwidth]{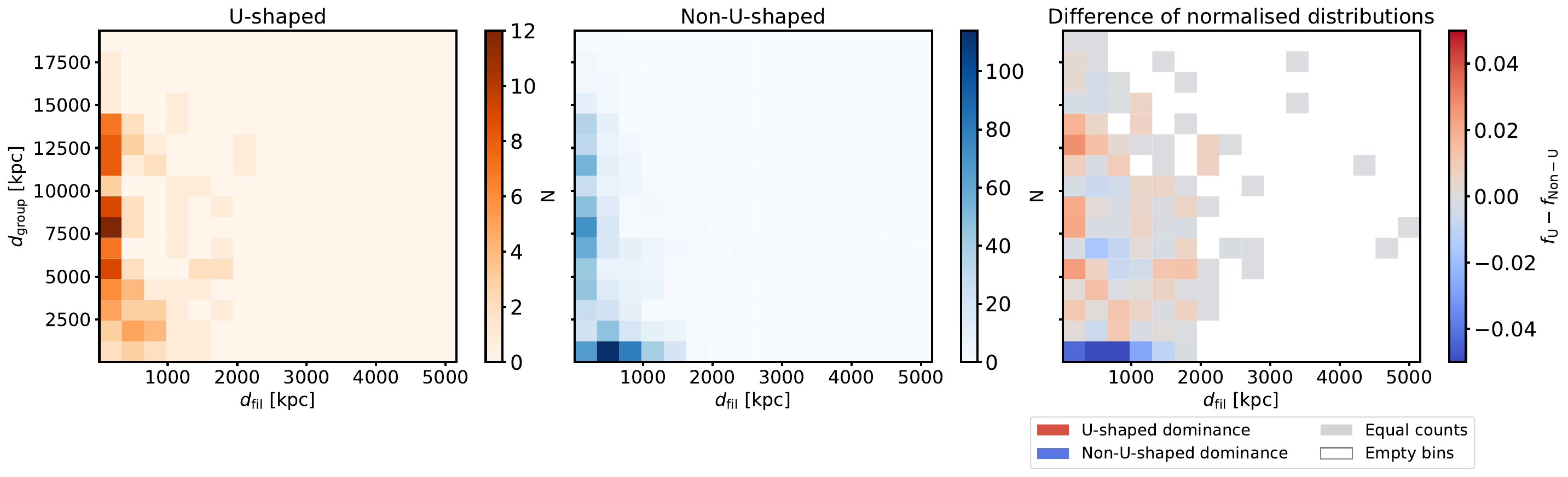}
      \caption{Distance of the galaxies with (left panel) and without (middle panel) a U-shaped profile to their nearest galaxy group, as a function of their distance to the nearest filament. Right panel: same as the left and middle panels, but considering the whole sample, colour-coded by the difference of the normalised distributions. We note that systems exhibiting a U-shaped feature in their profiles are close to filaments but far away from galaxy groups.}
         \label{fig:distgroups-filaments}
   \end{figure*}
   
To further characterise the environment of these galaxies, we computed their distance to the nearest filament ($d_{fil}$). To identify the filaments, we used the DisPerSE algorithm \citep{Sousbie2011}, which allows for a scale-free and parameter-free identification of the voids, walls, filaments, clusters and their configuration within the cosmic web, directly from the discrete distribution of particles (see Appendix \ref{appendixB} for details). 
We also computed the distance from each galaxy to its nearest group ($d_{group}$), defined as halos with $M_{200} \geq 10^{13} \, M_\odot$. In the left and middle panels of Fig.~\ref{fig:distgroups-filaments}, we show $d_{group}$ as a function of $d_{fil}$ for both galaxies with and without U-shaped age profiles, respectively. We note that galaxies 
with U-shaped profiles (left panel) lie typically near filaments (median $d_{fil} \approx 265$ kpc), while at the same time remaining at large distances from galaxy groups (median $d_{group} \approx 7903$ kpc). In contrast, galaxies without this type of profile (middle panel) can normally be found at larger distances from filaments, with a median distance of $\sim 351$ kpc, but closer to galaxy groups (median $d_{group} \approx 5861$ kpc). 
In the right panel of Fig. \ref{fig:distgroups-filaments}, we show this result for the whole sample, colour-coded by the difference of the normalised distributions, i.e. the fraction that one population occupies per bin minus the other ($f_U -f_{Non-U} = \frac{P_{U,~i }}{P_{U,~total}} - \frac{P_{Non-U,~i}}{P_{Non-U,~total}}$, where $P$ represents the population of each sample in the $i_{th}$ bin). Red colours indicate a dominance of systems with U-shaped age profiles. This confirms that these systems tend to be closer to filaments and farther from groups than galaxies without the U-shaped feature. 

   \begin{figure*}[ht!]
   \sidecaption
   \centering
    \includegraphics[width=12.95cm]{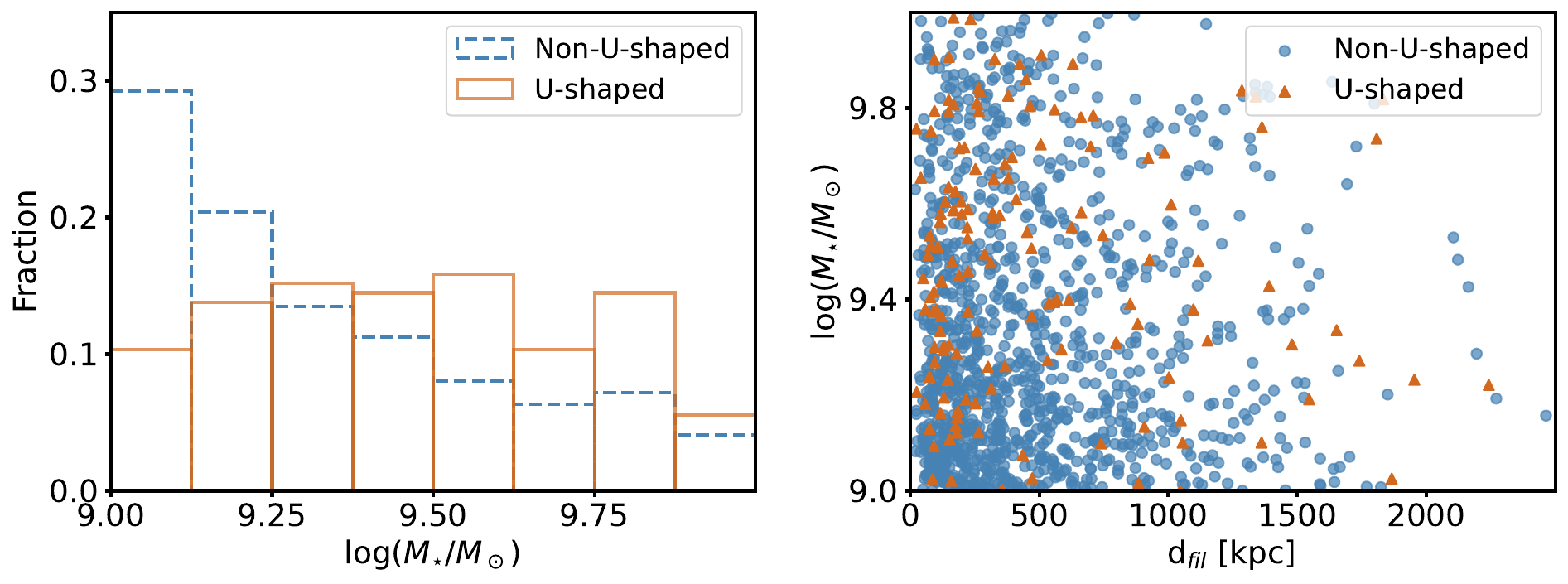}
      \caption{Left panel: stellar mass distribution of galaxies with a U-shaped age profile (solid) and without one (dashed). Right panel: stellar mass of galaxies with (triangles) and without (circles) U-shaped age profiles as a function of their distance to the closest cosmic filament. We note that those systems exhibiting this type of profile are closer to these structures.}
         \label{fig:distfilaments}
   \end{figure*}

As we mentioned in Sec.~\ref{subsec:tng50}, galaxies analysed in this work have stellar masses between $10^9$ and $10^{10} \, M_\odot$. In the left panel of Fig. \ref{fig:distfilaments}, we show the distribution of stellar masses of both samples of galaxies. Notably, galaxies with a U-shaped age profile are evenly distributed across the whole mass range, while they become increasingly dominant over non-U-shaped ones at higher stellar masses. This suggests that this feature in their profiles is relatively more common among the more massive galaxies, in agreement with the results of \cite{Tau2026} using the Auriga simulations. 

In the right panel of Fig.~\ref{fig:distfilaments}, we show the stellar mass of the galaxies in our sample as a function of their distance to the filaments. We find no correlation between the stellar mass of the systems exhibiting the U-shaped feature and their distance to the filaments, indicating that the more massive galaxies with this kind of profile are not the only ones close to these structures. We also find no correlation in our other sample: galaxies without a U-shaped age profile are, on average, farther away from the filaments, but this is the case for both more massive and less massive systems. 


\section{Discussion} \label{sec:discussion}

After analysing the distribution of low-mass galaxies with U-shaped radial age profiles in the previous section, it comes to our attention that these systems preferentially reside in low-mass halos and are usually central galaxies, as opposed to low-mass galaxies without this feature in their age profiles, which usually are satellite galaxies that reside within more massive halos. 
Additionally, as shown in Fig.~\ref{fig:distgroups-filaments}, galaxies with U-shaped age profiles reside mostly close to cosmic filaments. To confirm that this spatial preference is not driven by stellar mass bias, since more massive galaxies tend to reside closer to filaments \citep[e.g.][]{Malavasi2017}, we divided our sample into lower-mass ($\log(M_*/M_\odot) < 9.5$) and higher-mass ($\log(M_*/M_\odot) \geq 9.5$) subsamples (see Appendix \ref{appendixC} for further details). The trend remains consistently present in both mass regimes, demonstrating that the proximity to filaments is an intrinsic feature of galaxies exhibiting a U-shaped age profile rather than a mass-dependent effect.

Filamentary environments can shape these radial age profiles through several physical mechanisms that modulate a galaxy's gas reservoir. Filaments can induce starvation (or strangulation) by cutting off the accretion of fresh cold gas \citep[e.g.][]{Peng2015, AragonCalvo2019}, leading to a gradual decline in star formation. Furthermore, interactions with the intra-filament medium, such as cosmic web stripping \citep{Kraljic2018} or weak ram-pressure stripping, can remove gas from the outer parts while leaving the central regions intact, potentially favouring the rising of a U-shaped age profile. It is also worth noticing that the high frequency of merger events near filaments \citep{Dulcien2026} could impact the age profiles by directly affecting the galaxy's in situ star formation, and this in situ stellar material was shown by \citet{Tau2026} to be the primary component responsible for producing the U shape.

Regarding galaxies without U-shaped age profiles, we found that $\sim 47\%$ of this sample are central galaxies, indicating that neither satellite nor central evolutionary pathways can be neglected. In the case of satellites, a plausible explanation for why this status could inhibit the formation or survival of a U-shaped radial age profile is that environmental processes affecting them, such as ram-pressure stripping \citep[e.g.][]{Gunn1972} and tidal interactions, quench outer star formation or remove older halo stars, thereby altering the radial age distribution, which in low-mass systems usually translates into positive age gradients \citep[e.g.][]{Riggs2024}. Conversely, for central galaxies, smooth and continuous gas accretion from the surrounding intergalactic medium may independently prevent or erase U-shaped profiles. These systems can receive ongoing gas inflows across both their inner and outer regions \citep[e.g.][]{Sancisi2008, Putman2012}, sustaining star formation at large radii. This extends the spatial distribution of young stellar populations outwards, effectively washing out any potential age inversions, as found in Tau et al. (in preparation). 

Therefore, the gas content distribution in the environments in which low-mass galaxies with and without U-shaped age profiles reside plays a key role in the suppression and/or development of this feature, affecting their star formation history. This, alongside with the impact that galaxies' morphology may have on these results, will be further investigated in an upcoming paper. 



\section{Summary and conclusions} \label{sec:conclusions}

We have analysed a sample of $1301$ low-mass galaxies ($10^9 \, M_\odot \leq M_* \leq 10^{10} \, M_\odot$) using IllustrisTNG-50 simulations and found that 145 ($11.1\%$) of these systems have a U-shaped radial age profile. Most of the galaxies exhibiting this kind of profile ($\sim 72\%$) are central systems, and they typically reside in low-mass halos ($\sim 82\%$ of them in halos with $M_{200} < 10^{12} \, M_\odot$).

We found that the population of galaxies with U-shaped age profiles is located close to cosmic filaments, being $\sim 265$ kpc the median distance to these structures, while at the same time it is located at large distances from galaxy groups (median $d_{group} \approx 7903$ kpc). Hence, filaments could help develop the U-shaped feature in the age profiles of these systems due to the effect they have on their gas reservoirs through starvation (or strangulation) processes. Simultaneously, galaxy groups could have the opposite effect on galaxies, preventing the U shape feature to emerge in their age profiles. 

In contrast, galaxies without U-shaped age profiles tend to be located farther away from filamentary structures, with a median distance of $\sim 351$ kpc, but closer to galaxy groups (median $d_{group} \approx 5861$ kpc). In this case, smooth gas accretion from the intergalactic medium may deposit gas material in the outer regions of these galaxies, facilitating the formation of stars at large radii, and thus washing out any age reversal that otherwise could have been detected in the outer parts.


\bibliography{Bibliography}

\begin{appendix}
\nolinenumbers


\section{Acknowledgements}

\begin{acknowledgements}
      EAT would like to thank Lionel Messi for all the joy he brought to the Argentinians. EAT, DP, RBS, RS, and SVL acknowledge support from the Millennium Science Initiative Program NCN2024\_112. DP acknowledges support from the Agencia Nacional de Investigación y Desarrollo (ANID) through Basal project FB210003 and FONDECYT Regular projects 1241426. RBS acknowledges financial support from the Agencia Nacional de Investigación y Desarrollo (ANID)/Subdirección de Capital Humano/Doctorado Nacional/2023-21231017. AM acknowledges support from the FONDECYT Regular grant 1212046,  from the ANID BASAL project FB210003, and funding from the Max Planck Society through a ``PartnerGroup" grant as well as the HORIZON-MSCA-2021-SE-01 Research and Innovation Programme under the Marie Sklodowska-Curie grant agreement number 101086388. SVL acknowledges financial support from ANID/Fondo ALMA 2024/31240070.
\end{acknowledgements}

\section{Sample selection} \label{appendix}

To ensure profile purity, we excluded low-mass galaxies located in close proximity to relatively massive companion systems (i.e. $M_{*sat} \geq \frac{1}{100} \times M_{*host}$). In IllustrisTNG, the \textsc{Subfind} algorithm can misassign stellar particles between closely interacting or overlapping galaxies. Without this isolation criterion, stellar contamination from a nearby companion could be misattributed to the host system, producing an artificial outer age contrast and falsely generating a U-shaped radial profile.

To confirm that our conclusions are insensitive to the choice of spatial and mass thresholds, we tested several alternative masking criteria: (i) removing all close neighbours within $2 \,R_{hm}$ regardless of mass, (ii) restricting the exclusion zone to within $2 \,R_{hm}$ only for relatively massive companions to prevent contamination in the inner profile, and (iii) masking relatively massive companions within both $2 \,R_{hm}$ and $4\text{--}8 \, R_{hm}$. Across all tested variants, our primary findings regarding the spatial distribution and environmental preferences of galaxies with U-shaped age profiles remain unchanged. We therefore chose the stricter criterion that excluded relatively massive companions within $8 \, R_{hm}$.

\section{Central versus satellite galaxies} \label{appendixA}

\begin{figure*}[ht!]
   \centering
    \includegraphics[width=2\columnwidth]{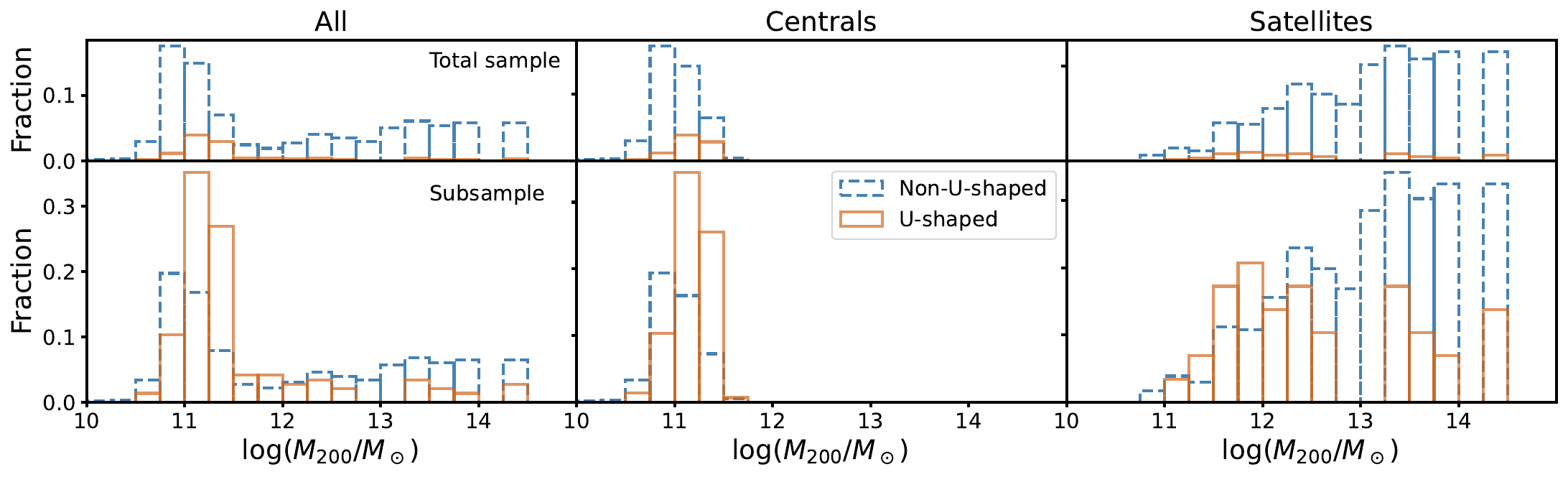}
      \caption{Environmental distribution of the galaxies of our sample. Left panels: distribution of the halo mass of the hosts of all galaxies with U-shaped age profiles (solid line) and all galaxies without these profiles (dashed lines). Middle panels: same as left panel but only for central galaxies in each subsample. Right panels: same as left panel but only for satellite galaxies in each subsample. The upper rows show these fractions computed regarding the whole sample of galaxies, whereas the lower panels show the fractions computed considering each subsample (i.e. galaxies with and without U-shaped age profiles).}
         \label{fig:centvssat}
   \end{figure*}

As we showed in Sec. \ref{sec:results}, low-mass galaxies with a U-shaped radial age profile mostly reside within low-mass halos ($M_{200} < 10^{12} \, M_\odot$). Here, we show that this is the case for both central and satellite galaxies exhibiting this feature. In Fig. \ref{fig:centvssat}, we present the halo mass distribution for all galaxies of our sample (left panel), as well as for central and satellite galaxies (middle and right panels, respectively). In the upper row, we show these results when considering the whole sample ($1301$ low-mass galaxies), whereas in the lower row we consider the fraction computed over each subsample (i.e. galaxies with a U-shaped feature and galaxies without it). 

We note that all central galaxies exhibiting a U-shaped age profile reside within halos of masses $M_{200} < 10^{12} \, M_\odot$. Similarly, satellite galaxies with this kind of profile also prefer lower-mass halos, given that we find the majority of them residing within halos of $M_{200} < 10^{12} \, M_\odot$, as opposed to satellite galaxies without this U-shaped feature, which tend to reside within more massive halos. Therefore, low-mass galaxies with a U-shaped age profile typically inhabit low-mass halos regardless of whether they are central or satellite galaxies.

\section{Identification of cosmic filaments using DisPerSE} \label{appendixB}

Filament identification was performed using the Discrete Persistence Structures Extractor \citep[DisPerSE\footnote{\url{https://www2.iap.fr/users/sousbie/disperse.html}};][]{Sousbie2011}, a topological technique based on discrete Morse theory and persistence theory. DisPerSE has been widely used to reconstruct the cosmic web in both numerical and observational works \citep[e.g.][]{Kuchner2020, Baier2025}. 

Briefly, DisPerSE takes the three-dimensional positions of halos as input and tessellates the simulation volume into cells, edges, and vertices. It estimates the underlying density field using the Delaunay Tessellation Field Estimator, which computes the density at each vertex of the Delaunay complex \citep[][]{Schaap2000}. To identify filaments and nodes, DisPerSE locates the critical points of the density field (maxima, minima, and saddle points) and connects them through its topological structure. The filaments are identified by computing small segments that connect saddle points to maxima (nodes), forming a continuous skeleton of the cosmic web. The significance of each connection is quantified by its persistence, defined as the density contrast between a pair of critical points. This quantity is typically expressed in units of the standard deviation ($\sigma$), with higher persistence thresholds allowing the removal of less significant, spurious structures.

In this work, filaments were identified in three dimensions using the halo positions across the entire TNG50 simulation box. The Delaunay tessellation was performed assuming smooth boundary conditions. Before extracting the filaments and critical points, each Delaunay cell was weighted by the mass of its corresponding halo, normalized by the mean halo mass across the simulation box. To extract the filaments, we adopted a persistence threshold of $6.5\sigma$ and a skeleton smoothing level (S) of $10$. We also explored a range of values for both parameters to assess their impact on the filamentary reconstruction, finding no significant differences in the main conclusions of this work. For instance, we used $S = 5, 8, 10$, and $20$, and found that galaxies with U-shaped profiles were always closer to the filaments than those systems without this feature regardless of its value, although the median distance to these structures could differ at most in $20\%$.

\section{Mass dependence test} \label{appendixC}

\begin{figure*}[ht!]
   \centering
    \includegraphics[width=2\columnwidth]{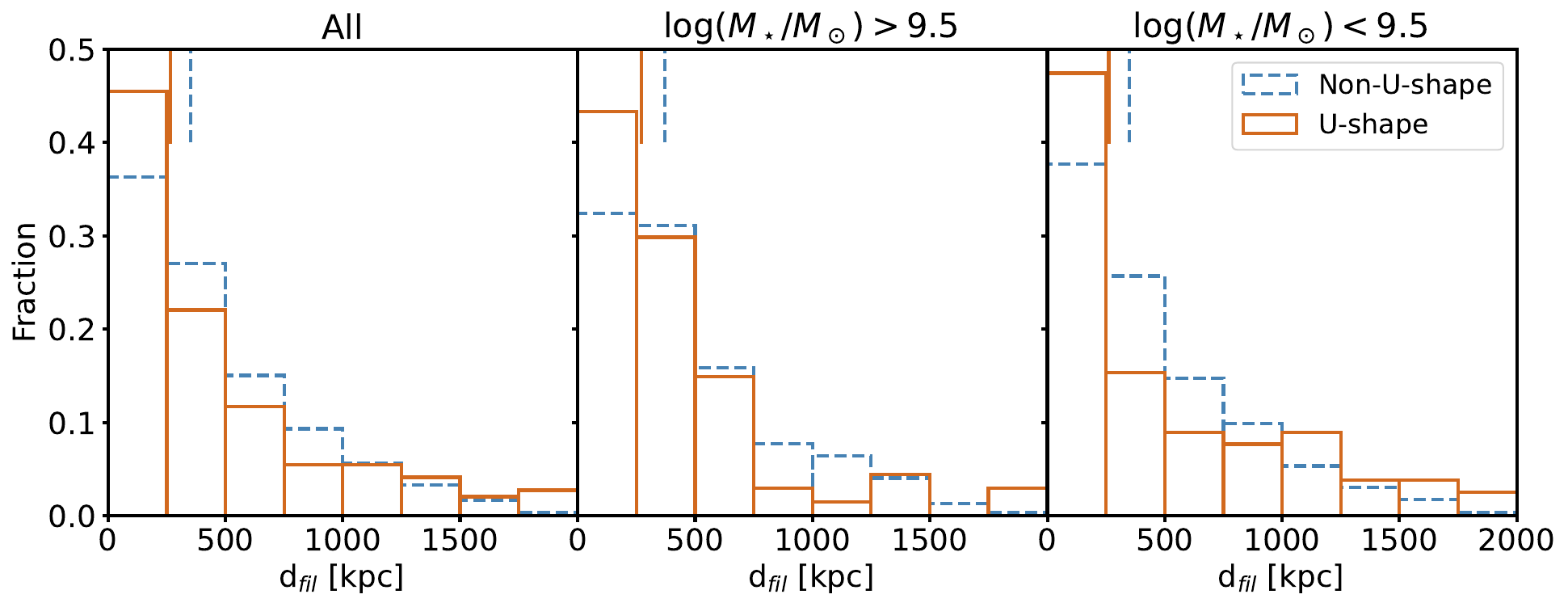}
      \caption{Distribution of the distance from the galaxies of our sample to their nearest filament. Left panel: distribution of the whole sample; middle panel: distribution of the higher-mass subsample; right panel: distribution of the lower-mass subsample. Solid lines represent the distribution of galaxies with U-shaped profiles, whereas dashed lines represent the distribution of those without this feature. The median distances to the filaments for galaxies exhibiting a U-shaped feature and galaxies without this profile are shown as upper solid and dashed vertical lines, respectively.}
         \label{fig:masstest}
   \end{figure*}

Since more massive galaxies tend to be closer to filaments than less massive ones, it is worth exploring whether the obtained relation between galaxies with U-shaped age profiles and their closeness to these structures is due to the fact that these systems dominate the high-mass end over galaxies without these type of profiles (see left panel of Fig. \ref{fig:distfilaments}). To do so, we divided our sample into lower-mass ($\log(M_*/M_\odot) < 9.5$) and higher-mass ($\log(M_*/M_\odot) \geq 9.5$) galaxies, and analysed their $d_{fil}$ distribution. These results are shown in Fig. \ref{fig:masstest}. We find that galaxies exhibiting a U-shaped age profile are always residing closer to filaments than those systems without this feature, regardless of their stellar mass. Hence, it is a fact that galaxies with U-shaped profiles reside close to filaments, rather a bias generated by a mass-dependence.

\end{appendix}
\end{document}